# Inductive sensing of magnetic microrobots under actuation by rotating magnetic fields

Michael G. Christiansen*[1], Lucien Stöcklin[1,2], Cameron Forbrigger[1], Shashaank Abhinav Venkatesh [1,3], Simone Schuerle*[1]

1 Department of Health Sciences and Technology, ETH Zurich
2 Department of Biosystems Science and Engineering, ETH Zurich
3 Department of Biomedical Engineering, National University of Singapore

*Michael G. Christiansen and *Simone Schuerle

**Email:** michael.christiansen@hest.ethz.ch, simone.schuerle@hest.ethz.ch

**Author Contributions:** MC and SS designed the research. MC, LS, and SV contributed to the iterative design and construction of the prototype apparatus. MC collected and analyzed data. CF characterized the amplifier and performed the theoretical sensitivity analysis. MC wrote the manuscript. All authors reviewed and edited the manuscript.

**Competing Interest Statement:** SS is a co-founder and member of the board of Magnebotix AG.

**This PDF file includes:**

>Main Text
>Figures 1 to 4
>References
>
>Supporting text
>Figures S1 to S6
>SI References

**Abstract**

The engineering space for magnetically manipulated biomedical microrobots is rapidly expanding. This includes synthetic, bioinspired, and biohybrid designs, some of which may eventually assume clinical roles aiding drug delivery or performing other therapeutic functions. Actuating these microrobots with rotating magnetic fields (RMFs) and the magnetic torques they exert offers the advantages of efficient mechanical energy transfer and scalable instrumentation. Nevertheless, closed-loop control still requires a complementary noninvasive imaging modality to reveal position and trajectory, such as ultrasound or x-rays, increasing complexity and posing a barrier to use. Here, we investigate the possibility of combining actuation and sensing via inductive detection of model microrobots under field magnitudes ranging from 100s of µT to 10s of mT rotating at 1 Hz to 100 Hz. A prototype apparatus accomplishes this using adjustment mechanisms for both phase and amplitude to finely balance sense and compensation coils, suppressing the background signal of the driving RMF by 90 dB. Rather than relying on frequency decomposition to analyze signals, we show that, for rotational actuation, phase decomposition is more appropriate. We demonstrate inductive detection of a micromagnet placed in distinct viscous environments using RMFs with fixed and time-varying frequencies. Finally, we show how magnetostatic gating fields can spatially isolate inductive signals from a micromagnet actuated by an RMF, with the resolution set by the relative magnitude of the gating field and the RMF. The concepts developed here lay a foundation for future closed-loop control schemes for magnetic microrobots based on simultaneous inductive sensing and actuation.

**Significance Statement**

Magnetic microrobots are anticipated to eventually be able to navigate the body and perform medical functions. While the use of magnetic fields to control their motion is well studied, getting real time feedback on their position and movement usually requires a parallel imaging technique. Here, we use magnetic fields to both sense and actuate microrobots at the same time, designing a setup that uses symmetry to isolate voltages induced by the stray fields of model microrobots. We consider how inductive signal processing must differ from the more familiar approach used in magnetic particle imaging. Moreover, we demonstrate how inductive signals can be isolated to points in space under rotational actuation when the rotating field is combined with a gating field.

**Main Text**

**Introduction**

Magnetic stimuli offer an appealing means to control medical microrobots by providing noninvasive access to locations deep within the body (1-4). Limitations in scaling up magnetic field gradients to the dimensions of human patients, particularly gradients sufficient to manipulate the smallest microrobots, have fueled interest in locomotion schemes that are instead driven by magnetic torques (4-6). Microrobots have, for instance, been engineered to swim through fluids in a corkscrew motion (7, 8) or tumble (9-11) or roll (12-15) along surfaces in response to comparatively scalable magnetic fields with magnitudes of mT rotating at frequencies of Hz to 100s of Hz. When applied to individual microrobots or swarms of microrobots (16), these rotating magnetic fields (RMFs) perform mechanical work through torque, transferring energy from the circuits driving the surrounding electromagnets to the microrobots and their surroundings (7, 8, 12, 16).

One of the longstanding challenges for the deployment of medical microrobots in relevant physiological situations like site-specific drug delivery is the implementation of closed loop control, which typically requires simultaneous imaging and actuation (17, 18). While efforts have been made toward using complementary imaging techniques based on ultrasound or x-rays, a currently underexplored advantage of using time-varying magnetic stimuli for actuation is that the dynamic response of the microrobots themselves can generate inductively detectable voltage signals (19).

This possibility for inductive detection is perhaps best appreciated through analogy to magnetic particle imaging (MPI), a technique in which voltage signals generated by the time-changing magnetization of magnetic nanomaterials are spatially selected with a gating field (GF) to reconstruct images (20, 21). Despite their similar underlying sensing principle, there are substantial and challenging technical differences between MPI and inductive detection with RMFs (Fig. *1*). MPI employs much higher frequencies (kHz) with proportionally higher induced voltages, suggesting lower detection sensitivity for RMFs with a frequency below 100 Hz. Indeed the magnetic stimuli required for MPI and RMF actuation are so distinct that it is possible to employ them sequentially on superparamagnetic swimming robots, albeit with lag during imaging sequences (22). Here, we instead emphasize the possibility of inductively detecting the response of microrobots under the conditions used to actuate them. One important consequence of using RMFs rather than alternating magnetic fields to drive magnetization response is the elimination of the periodic saturation of magnetic material that leads to readily separable higher harmonic signal contributions (Fig. *1*).

In this manuscript, we develop methodology and demonstrate a prototype for inductively detecting the response of individual microscale magnets to RMFs, which serve as model microrobots. The crucial task of cancelling background signal from the RMF is accomplished through two separate mechanisms: a potentiometer that balances in-phase background signal and a set of orthogonal phase-adjusting loops alongside the compensation coil that can be selectively incorporated into the circuit to balance the out-of-phase background signal. Rather than relying on frequency decomposition to identify harmonic contributions from magnetic material as in MPI (21), we instead show that for inductive detection with RMFs, it is more relevant to consider phase decomposition. The out-of-phase component arising from the response of the microrobots is especially informative, in part because it is insensitive to unwanted diamagnetic or paramagnetic contributions, and because its magnitude can be related to irreversible work done on the microrobot(s). The expected relationship between design parameters of both the microrobotic component and the inductive detection apparatus is considered and elucidated. Finally, we show that inductive detection with RMFs is compatible with magnetostatic GFs, with the resolution of the zero-point set by the relative field magnitude of the RMF and the selection field gradient. The techniques and concepts developed here represent a step toward the seamless integration of actuation and sensing of magnetic microrobots with scalable time varying magnetic fields.

**Results and Discussion**

**Phase decomposition enables inductive detection with low frequency rotating magnetic fields**

Like MPI or magnetic particle spectroscopy (MPS), inductive detection of microrobots actuated by RMFs involves the isolation of induced voltage signals produced by magnetically responsive materials surrounded by media with distinct and substantially weaker magnetic properties (21). In MPI and MPS, an alternating magnetic field is applied to drive periodic magnetic saturation, leading to readily separable higher order harmonic contributions to the inductively detected signal. By contrast, the use of an RMF, which remains constant in magnitude as it sweeps through a plane of rotation, precludes the possibility for signal isolation based on periodic saturation (Fig. *1*). Instead, under the conditions of steady state rotation that are often applicable to the locomotion of microrobots, the induced signal is purely sinusoidal, with a magnitude proportional to the moment of the microrobot and a phase shift attributable to the torque balance between viscous drag and magnetic torque (Fig. *1*).

In the limit of zero drag, a microrobot rotates perfectly in-phase with an RMF, experiencing no instantaneous magnetic torque as it is carried along in its constant rotational velocity by its angular momentum. Under realistic conditions, the out-of-phase component of the magnetization corresponds to the magnetic torque being continuously applied to counteract drag or friction. While frictional and drag forces and torques are dissipative in the sense that they do irreversible work, magnetic microrobots actuated by RMFs are typically designed in such a way that the resulting convective flows or boundary interactions give rise to translational motion or surface walking (7, 9, 13). The out-of-phase component of the magnetization in these conditions is therefore useful to detect in part because its magnitude relates directly to the energy transfer from the field to the microrobot, a quantity that is often desirable to maximize. Another motivation for detecting the out-of-phase component of the magnetization is that this part of the signal has no analogous contribution from the paramagnetic or diamagnetic response of surrounding materials, and thus might play a role similar to the higher frequency harmonics in MPI and MPS that arise uniquely from the behavior of the magnetic material (Fig. *1*).

Phase decomposition of the detected signal from a microrobot can be expressed mathematically in several different ways. One computationally efficient method might be to perform a fast Fourier transform (21) and consider the real and imaginary parts at the frequency of the RMF. Here, to facilitate comparison with the techniques we describe later for signals with time-varying frequency, we choose instead to formulate the phase decomposition in terms of a Fourier integral with its basis limited to sines and cosines at the same frequency as the RMF. For a microrobot exposed to an RMF, the induced signal from the x projection of the magnetization, $V_{m_x}$, can be expected to vary as a function of time $t$ as follows:

$$V_{m_x}(t) = V_\parallel \sin \omega t + V_\perp \cos \omega t \qquad [1]$$

Here, $\omega$ is the angular frequency and $V_\parallel$ and $V_\perp$ are constant coefficients that can be determined through integration over one or more periods. Specifically,

$$V_\parallel = \frac{\omega}{\pi} \int_{t_i}^{t_f} V_{m_x}(t) \sin(\omega t)\, dt \qquad [2]$$

where the limits of integration $t_i$ and $t_f$ are any integer number of periods apart. Analogously,

$$V_\perp = \frac{\omega}{\pi} \int_{t_i}^{t_f} V_{m_x}(t) \cos(\omega t)\, dt \qquad [3]$$

Provided that 1) the signal induced by the x projection of the RMF, $V_{H_x}$, is in-phase with $\sin(\omega t)$, and that 2) the symmetry of the detection coil geometry ensures the true phase lag of the microrobot is undistorted in the detected signal, the coefficients $V_\parallel$ and $V_\perp$ describe the magnitude of the in-phase and out-of-phase contributions to $V_{m_x}(t)$, respectively (19). Due to the fundamental dependence of induced signals on the time derivative of magnetic flux density, the scale of voltage signals induced at different fixed frequencies are proportional to $\omega$. This fact, taken with the coincidentally similar relationship between mechanical power transfer and frequency, implies that $V_\perp$ is directly proportional to the rate of irreversible work done on the microrobot. Alternatively, dividing $V_\parallel$ and $V_\perp$ by $\omega$ or frequency can enable comparison between normalized signals that are proportional to vector projections of the magnetic moment for measurements performed at different frequencies. These signals can be related back to the measured voltage by some factor of proportionality that depends on the geometry of the coils and amplification within the detection circuit, which could be determined using a sample with known properties similarly to calibration of vibrating sample magnetometers (23). With this normalization, $V_\perp$ can be interpreted as proportional to the torque continuously applied by the field or, equivalently, proportional to the irreversible work done per cycle of the RMF.

**Background cancellation with rotating fields requires mechanisms for adjusting in-phase and out-of-phase contributions**

While it would, in principle, be possible to use various alternative types of magnetic field sensors to detect stray field contributions from magnetic microrobots in an RMF, inductive detection is a favorable approach in practice because it permits designs for which the voltages induced by the RMF can be physically cancelled within the coil itself prior to signal amplification. Often the measurement capabilities of such a setup are not limited directly by the scale of the detected signal voltage, but rather by the extent to which the contribution of the microrobot is sufficiently distinguishable from whatever residual uncompensated signal remains from the driving RMF. While coil geometries that achieve perfect cancellation can be readily conceived abstractly, a suitable design for a real prototype must be combined with effective adjustment mechanisms that allow for fine tuning of the background cancellation.

In the detection apparatus constructed for this work, the RMF was supplied by a two-phase set of armature coils arranged to generate an RMF within a cylindrical working volume in the plane perpendicular to the axis of the cylinder. At the center of the cylinder, on a 3D printed form, two sets of pickup coils were situated orthogonally to detect induced voltages from $H_x$ (blue) and $H_y$ (yellow) with a total of 1000 turns each (Fig. *2A*). These are required to monitor and record the time-varying components of fields applied to the working volume, for instance to ensure circular rotation and determine the phase shift between the RMF and the microrobot. Above these field pickups, a pair of coils with a total of 5000 turns of 50 µm magnet wire was situated symmetrically on either side of the detection zone (orange, Fig. *2A*). This set of coils is mirrored across the central plane of the cylinder by another set of coils with 5000 turns that acts as the compensation coil (teal, Fig. *2A*). These coils are connected in series as indicated in the schematic in Fig. *2A*.

Background cancellation requires balancing the voltages between the sense coils and compensation coils as closely as possible in the absence of a microrobot. Because the driving field is an RMF, this means both the balance of their magnitude and their phase need to be finely adjusted (19). For in-phase amplitude adjustment, the signal was taken from the wiper of a 10-turn potentiometer before amplification, allowing for fine manual adjustment of the voltage divider (Fig. *2A*, right inset). Multiple strategies can contribute to balancing the phase between these two coils.

For rough cancellation, one strategy is to adjust the tension in the screws at top and bottom of the 3D printed coil holder to minutely adjust its precise angle with respect to the driving coil. For fine tuning the phase cancellation in a systematic manner, a set of concentrically arranged orthogonal loops of varying area (light blue, Fig. *2A*) were formed and selectively incorporated into the sense and compensation coil circuit with double-pole-double-throw switches. The effective area of these loops varied by approximately a factor of 4 over 4 levels of adjustment, such that each set of 5 switches at the same level could span the out-of-phase signal contribution of a single switch at the next coarsest level. This resulted in an array of 20 switches to provide adequate range and precision for the adjustment. The polarity of these loops could be changed with another switch (Fig. *2A*), such that this scheme could introduce phase components either leading or lagging as needed. A typical adjustment involved iterative manipulation of the potentiometer and switches, incorporating the loops with the largest area first and using the smaller loops to attain progressively finer cancellation (Fig. *2A* right inset).

As a model microrobot, cylindrical neodymium iron boron micromagnets with 300 μm diameter and 500 μm length were employed in suspensions of water and pure glycerol (Fig. *2B*). Despite their lack of immediate biomedical function, these micromagnets were adopted as a convenient model system that recapitulates the basic physics of rotating magnetic bodies in viscous media outlined in the previous section. To observe signals from rotation of the micromagnets at the highest frequencies, the amplification of the two-stage amplifier had to be reduced from an overall gain of approximately 120 dB to 70 dB (*SI Appendix* Fig. S1). A representative example of an averaged raw signal is shown in Fig. *2C*, along with the residual background remaining when a vial without a micromagnet is placed back into the coil. A background subtracted signal is shown in Fig. *2D*, alongside the voltage signals induced in the pickup coils for the field. The scale of this residual signal indicates that the cancellation scheme suppresses the induced signal from the driving field by 90 dB (*SI Appendix* text). By applying the mathematical techniques detailed in the previous sections, a fit was determined in terms of the coefficients $V_\parallel$ and $V_\perp$, and the resulting curve, which contains no contributions from 50 Hz noise, is shown as well.

Similar signals were collected and processed over a range of discrete frequency values for a micromagnet in glycerol and a micromagnet in water and normalized by frequency (Fig. *2E*). The results generally recapitulate the expected physical behavior for these systems. At low frequencies, both micromagnets rotate almost entirely in-phase with the field, with the out-of-phase component of the signal (and thus magnetic torque) increasing with frequency. The micromagnet in glycerol requires larger torques to maintain rotation at all frequencies, and its out-of-phase component reaches comparable values to the in-phase component at a lower frequency than the micromagnet in water.

These observed behaviors comport well with analogous experiments performed using the well-established technique of rotational magnetic spectroscopy (24), which involves observing microscopic magnetic objects with brightfield microscopy as they rotate in RMFs of varying frequency and magnitude. The equations of motion in viscoelastic media have been described (25), including the regime of asynchronous motion above the critical frequency at which the micromagnet can no longer synchronously keep pace with the RMF. A somewhat analogous technique based on inductive detection called rotational drift spectroscopy also exists, which applies RMFs at 10s of kHz to magnetic nanomaterials to observe shifts in hydrodynamic diameter in response to agglomeration or binding to analytes (26, 27). One key difference in the present work is the frequency range (1 Hz to 100 Hz) set by the locomotion requirements for microrobots. This explains the need for coils with as many turns as possible and the emphasis on finding suitable mechanisms for finely tuned background cancellation.

Given the greatly reduced operational frequency required for simultaneous microrobotic actuation and detection, it is useful to consider the ultimate detection limits of this prototype and similar setups. A theoretical sensitivity analysis was performed by approximating the detection coils here as projected partial spherical surfaces (Fig. *2F*, *SI Appendix,* text and Fig. S2). The resulting scale of the voltage signals are plotted in terms of a quantity on one axis called the "agent parameter" that is determined by characteristics of the microrobot, and a quantity on the other axis called the "coil parameter" that depends on aspects of the inductive detection apparatus. In terms of the how agent properties influence sensitivity, the detected voltage is proportional to the product of the magnetic moment of the microrobot and its preferred frequency of operation. If the properties of the microrobot are fixed and only aspects of the detection apparatus may vary, signals are proportional to the product of the number of turns and linear gain of the amplifier, and inversely proportional to the linear dimensions of the coil geometry. The conditions explored experimentally in this work are mapped onto this space with the orange rectangle (Fig. *2F*). Our work with this system suggests that it would be readily feasible to increase the gain of the amplifier and the number of turns for more sensitive detection (*SI Appendix*, Fig. S1). We note, however, that this analysis does not account for other factors that may limit performance. These include the presence of a 50 Hz noise floor, thermal drift of the resistance balance between the sense and compensation coils, limits to the sensitivity of background cancellation, and the possibility for dielectric breakdown in the limit of coils with extreme numbers of turns that produce excessive internal voltages. These issues are all arguably addressable by design choices and therefore the scale of the sensed voltage represents a more fundamental constraint.

**Phase decomposition can be extended to rotating fields with time-varying frequencies**

While using inductive detection to observe the response of a model microrobot to RMFs with discrete fixed frequencies is useful for capturing steady state response, a more general and broadly applicable case is that of dynamic response to driving fields with time-varying frequency. In a manner somewhat similar to variable frequency drives for industrial motors (28), an analogous technique can readily be envisioned for closed loop control of microrobots with RMF frequency varied to optimize their inductively detected phase lag response. To build toward this use case, the phase decomposition technique shown in Eq. **1-3** must be extended to the case of $\omega = \omega(t)$. In such a case, $V_\parallel$ and $V_\perp$ can be treated as time varying functions rather than constant coefficients, giving the following assumed functional form for $V_{m_x}(t)$:

$$V_{m_x}(t) = V_\parallel(t) \sin[\omega(t)t] + V_\perp(t) \cos[\omega(t)t] \qquad [4]$$

If $V_\parallel(t)$ and $V_\perp(t)$ vary slowly compared to the oscillation of the field and can be treated as approximately constant over one period, then it is possible to find approximate expressions for them at a time $t_0$.

$$V_\parallel(t_0) \approx \frac{2 \int_{t_i}^{t_f} W(t) V_{m_x}(t) \sin[\omega(t)t]\, dt}{\omega(t_f) t_f - \omega(t_i) t_i} \qquad [5]$$

And

$$V_\perp(t_0) \approx \frac{2 \int_{t_i}^{t_f} W(t) V_{m_x}(t) \cos[\omega(t)t]\, dt}{\omega(t_f) t_f - \omega(t_i) t_i} \qquad [6]$$

Here, $W(t)$ is a weighting function, defined in terms of $\omega(t)$.

$$W(t) \equiv \frac{d}{dt}[\omega(t)t] \qquad [7]$$

The limits of integration must account for the shifting frequency to describe a single period, such that $t_i$ and $t_f$ are functionally related in terms of $\omega(t)$.

$$\omega(t_f)t_f = \omega(t_i)t_i + 2\pi \qquad [8]$$

The midpoint of $t_i$ and $t_f$ can be taken as the time $t_0$ corresponding to the approximate instantaneous value of the coefficient functions.

To investigate the possibility of processing signals in this manner, we exposed the same micromagnets from the previous section to RMFs of constant magnitude with time-varying frequencies. Specifically, we made use of a rapid linear frequency sweep from 10 Hz to 110 Hz (Fig. *3A*) and a slower quadratic frequency sweep from approximately 0.6 Hz to 26 Hz (Fig. *3B*). For the linear sweep, the driving field advanced through the full frequency range in less than a second while its amplitude was kept constant (Fig. *3C*). The relatively wide frequency range of the linear sweep allowed the effect of the same driving conditions to be tested on nominally identical micromagnets in two different viscous environments, glycerol and water. The quadratic sweep was chosen to show that the methods developed here are also applicable to nonlinear frequency-vs-time functions. In theory, any frequency function could be chosen, provided the assumption of slowly-varying $V_\parallel$ and $V_\perp$ is reasonable. For the quadratic sweep, the frequency increased quadratically over five seconds, and the sweep was performed at two distinct amplitudes for the same microrobot in glycerol (Fig. *3D*). Frequency normalized, background subtracted curves from inductive detection of the microrobot are shown in Fig. *3E* and Fig. *3F*. The magnitude of the signal can be observed to drop off rapidly with increasing frequency in the case of the micromagnet in glycerol under a linear sweep, and the signal persists after the magnitude of the field has fallen in the case of the micromagnet in water. Otherwise, by inspection, many of these frequency normalized curves appear similar and phase decomposition analysis is needed to extract additional information.

Phase decompositions of the inductively detected signals from the linear sweep are shown in Fig. *3G*, and for the quadratic sweep in Fig. *3H*, with instantaneous values for $V_\parallel(t)$ and $V_\perp(t)$ shown instead as a function of frequency, $V_\parallel(f)$ and $V_\perp(f)$, based on the known time dependence of the frequency. Further details about the signal processing procedure can be found in the text of the *SI Appendix*. In the case of microrobots under conditions of different viscosity, phase decomposition reveals response curves that appear to be functionally similar but shifted to lower frequencies for the microrobot in glycerol due to the substantially greater viscous drag. If the goal were to maximize $V_\perp$ to optimize mechanical energy transfer based on the environment of the microrobot, a much higher frequency would obviously be needed in water than in glycerol. Curves like these could offer a basis for selecting or testing the optimality of drive frequency in real time as a microrobot encounters differing local conditions. Similarly, Fig. *3H* shows how phase decomposition reveals changes in frequency dependent response as a function of the magnitude of the RMF, which determines the corresponding scale of available magnetic torque. Notably, the crossover frequency between the in-phase and out-of-phase component of the signal can be seen to increase approximately proportionally to the RMF magnitude, as expected. At 11.1 mT, the crossover frequency was 12.2 Hz and at 20.7 mT the crossover frequency was 20.7 Hz. Wide or rapid frequency sweeps like the ones shown are most useful in the context of characterization, and for the purpose of closed loop control, the frequency would likely be varied less rapidly and over a smaller range of values.

**Magnetostatic gating fields allow for spatially restricted rotational actuation and inductive detection**

Previous work has indicated the possibility of using a magnetostatic GF to spatially isolate the delivery of magnetic torque to individual microrobots or magnetic torque density to diffuse assemblies of microrobots or magnetic material (19, 29-33). Mechanistically, by combining RMFs and magnetostatic GFs, the rotational character of the field is preserved only in the field free region and suppressed at points away from it where the magnitude of the superimposed GF becomes comparable to or higher than the RMF (29, 30). As discussed in a previous section, the out-of-phase component of the inductive signal detected from microrobots in an RMF reflects the mechanical work being done by magnetic torque. Together, these facts led us to hypothesize that inductive signals produced by microrobots in response to RMFs should also be spatially isolated by a magnetostatic GF, in a manner analogous to MPI. This would imply that, in addition to the possibilities for spatially selective rotational actuation that GFs offer, the inductive signals produced in them might also reveal positional information about the actuated microrobots.

To test this hypothesis, we constructed an array of permanent magnets capable of fully enclosing the inductive detection setup described in previous sections (Fig. *4A*, *SI Appendix,* Fig. S4). This array was based on a "magic sphere" geometry—a modified azimuthal revolution of a second order Halbach cylinder with a central field free region (34, 35). In this case, as with a similar array that we built previously for torque density focusing (29), a 3D printed support structure held ferrite ceramic permanent magnets (grade Y35) that were stacked to form uniformly magnetized cubes with an approximately 2 cm side length. On the inside of the partial sphere, around its equator, neodymium iron boron magnets were interleaved with the ferrite magnets to improve the resolution of the GF along the x and y axes. The results of a finite element model predicting the magnetic field generated by an axisymmetric approximation of this configuration of permanent magnets is shown in Fig. *4B*. This is best interpreted as an upper bound of the expected field under the assumption of a perfect circumferential packing factor. This geometry is expected to produce its strongest gradient in the z direction, and indeed the largest gradient observed (1.1 T/m) was measured along the z axis using a Hall probe controlled by a micro positioner (Fig. *4C*, *SI Appendix,* Fig. S4). The gradients observed along the x axis (0.8 T/m) and along the y axis (0.4 T/m) were unequal due to a column of omitted magnets along each side of the seam where the two halves of the sphere met. Although an isotropic zero point would likely be desirable for imaging or inferring position, in this case, the ellipsoidal field free region was useful because it permitted measurements of inductive signal at known positions along different magnetostatic gradients without reconfiguring the setup.

As a model microrobot confined to a geometrically isotropic environment, a micromagnet nominally identical to the ones employed in the previous sections was placed in a 3D printed 3 mm spherical cavity filled with glycerol and sealed with a polycarbonate sheet (Fig. *4D*). The sample chamber was affixed to an arm mounted on the micro positioner (*SI Appendix,* Fig. S4), which could also be used separately to position a three axis Hall probe in the same working space. Prior to inductive measurements of the microrobot, the time dependent net field arising from the superposition of the RMF and the GF was measured at positions along the principal axes of the GF, with several representative examples shown in Fig. *4D*. For points along the z axis displaced from the center, the net field is observed to undergo precession about the z axis (Fig. *4D* i). At the zero point, the field rotates in the xy plane with a magnetostatic contribution less than 0.1 mT (Fig. *4D* ii). At points along the x or y axis, the magnetostatic contribution occurs entirely within the xy plane, similarly suppressing the rotational character of the field (Fig. *4D* iii). Away from the principal axes, behavior intermediate between these special cases can be expected.

For the frequencies selected for this experiment, RMF magnitudes were determined empirically that maximized the out-of-phase component of the inductively detected signal when the sample chamber was placed at the zero point (*SI Appendix*, Fig. S5). Because the RMF magnitude was varied (1.8 mT at 11 Hz, 0.89 mT at 5.5 Hz, and 0.49 mT at 2.75 Hz), the effective resolution of the zero point could also be anticipated to change, with increasing spatial selectivity provided by the same GF as the RMF magnitude decreases (29). As expected, the high gradient of the GF along the z axis resulted in the highest observed spatial selectivity, with the out-of-phase inductive signal from the microrobot dropping to half its peak value within millimeters of the zero point (Fig. *4E*). Along the weakest gradient of the zero-point, corresponding to the y axis, spatial selection was also observed, though it was weaker (Fig. *4F*). For the highest RMF magnitudes tested, a plateau is observed in the immediate vicinity of the zero point along the y axis, behavior that vanishes as the RMF magnitude is decreased. To rule out the role of position-dependent inductive coupling to the sense coil as the origin of the spatial variation of these signals, the response of a magnet in glycerol was tested as a function of position in the absence of the GF (*SI Appendix*, Fig. S6). Although the inductive signal was found to vary with position, as expected, the spatial dependence is weaker than to the signal dropoffs observed with the selection field.

If the normalized out-of-phase signal datapoints from the positive displacements of Fig. *4E* and Fig. *4F* are replotted as a function of the relative magnitude of the GF to the RMF, the general behavior appears to converge toward a single curve (Fig. *4G*). This is consistent with our previous finding that the spatial resolution achieved when combining a rotating magnetic field and a magnetostatic GF is set primarily by the parameters of these two fields (29), a result that is intuitive in the context of vector superposition (Fig. *4D*). The fact that resolution is set primarily by field parameters is an important difference from MPI, where material properties play a key role in setting spatial resolution (21).

**Conclusion and Outlook**

Actuating microrobots with time-varying magnetic fields strongly suggests an opportunity to detect their response via induced voltage signals. While the stray fields of microrobots could, in principle, be observed directly by a magnetic field sensor with an exceptionally high sensitivity and a correspondingly high dynamic range, the inherent advantage of inductive detection is the possibility it creates for effective isolation of the signal of the microrobot. This isolation is accomplished primarily through geometric symmetry, requiring only simple components and a robust strategy for fine tuning cancellation of the signal from the driving field. In this work, we have designed prototype instrumentation and demonstrated signal processing for simultaneous inductive detection and actuation of a model microrobot with low frequency rotating magnetic fields. Moreover, we showed that this form of inductive detection can be combined with a magnetostatic GF for restricting actuation and the inductive signal it generates to a single point. Our findings suggest that inductive detection of microrobots could serve as a basis for closed-loop control, whether directed toward optimizing mechanical energy transfer or controlling microrobot motion.

For the technical potential of inductive detection of microrobots to be more fully realized, further development is required. For one, similar techniques should be adapted for application to smaller microrobots such as bacteria-inspired helical swimmers (6), magnetically responsive bacteria (19), or surface microrollers (14). The micromagnet picked here as a model microrobot does adequately reflect the basic physics of these systems, but even smaller microrobots are more likely to be deployed in a medical context. Our sensitivity analysis indicated that induced voltage signals are inversely proportional to linear dimensions of pickup coils, so detecting smaller microrobots in realistic working volumes will require increasing the number of turns, boosting amplification, and shielding setups from ambient 50 Hz noise. The cancellation techniques that were demonstrated in this work can also be further refined, especially through automation, since here they were implemented manually. Ultimately, the form factors of greatest interest for real applications could be single-sided (36, 37), i.e. based on detection coils that do not enclose the working volume, and executing this in practice would require studying how pickup coil geometry influences phase and magnitude of the induced signal. In this respect, the spatial isolation provided by the GF may also offer a practical path, since an integrated array of permanent magnets in a single sided design could isolate inductive signal generation to a point or line with known inductive coupling behavior.

Several different forms of noninvasive energy transfer and actuation are available to control medical microrobots, including ultrasound, light, and magnetic fields. Each of these has its own unique features and drawbacks. Up to this point, one of the major limitations of magnetic microrobot control has been the need for complementary imaging, which remains an outstanding challenge. The realization of simultaneous magnetic actuation and inductive detection of microrobots with time varying magnetic fields could represent a step toward feasible real-time closed-loop control.

**Materials and Methods**

**Construction of the inductive sensing setup.** Custom designed coil holders were 3D printed with a resin-based digital light processing printer and assembled with epoxy and plastic hardware. STL files can be made available upon request. Sense and compensation coils were wound by hand using enameled magnet wire with 50 µm diameter, 100 µm diameter magnet wire was used for the x and y field sensing coils, and 200 µm diameter magnet wire was used to wind the phase adjustment loops. The RMF coils were also wound by hand with 300 turns of magnet wire of 0.9 mm diameter, symmetrically alternating between the phases in groups of ten turns to maintain overall symmetry. Signals were generated with a 2-channel arbitrary function generator (Keysight EDU33212A) and fed to a two-channel class AB power amplifier with a rated maximum output power of 80 W per channel (Crown D-150A). The power amplifier was situated across the room from the setup to reduce 50 Hz noise. Pickup coils for sensing $H_x$ and $H_y$ were measured directly with an oscilloscope (Keysight DSOX2004A). The output from the sense and compensation coil pair was wired in series with the phase loop switchbox, which had one switch to determining sign of the out-of-phase contribution and could selectively incorporate up to 20 separate loops with 4 different magnitudes. Finally, the signal was taken from the wiper of a potentiometer and underwent two-stage signal amplification. Additional information on the amplifier design is provided in *SI Appendix*, Fig. S1.

**Phase decomposition at fixed frequencies.** Input signals were tuned manually to provide a constant target RMF magnitude and an appropriate quarter period phase shift between the $V_{H_x}$ and $V_{H_y}$ signals. The factor relating the induced voltage from the field probes to the RMF magnitude was found through empirical calibration at a single frequency via simultaneous measurements with a 3 axis Hall probe (MetroLab THM1176). The micromagnet (SM Magnetics, Cyl0003-50) was placed in pure water or pure glycerol in a standard 5 mm NMR tube, and blank samples containing only glycerol or water were similarly prepared. At each frequency, 5 separate measurements of the voltage signal were performed, with blank measurements performed before and after each trial and internal averaging on the oscilloscope set to 64. Background cancellation was fine-tuned prior to each sequence of measurements but left untouched between measurements of blanks and trials. Mathematica was used to analyze the resulting data. In brief, for fixed frequency measurements, a sinusoidal function was fitted to $V_{H_x}$ to determine the phase shift needed for the basis functions in Eq. 2 and 3, and the coefficients $V_\parallel$ and $V_\perp$ were found through numerical integration with $V_{m_x}$. For each trial, the background subtracted coefficients were determined by subtracting the mean of the blank measurements immediately before and after the trial.

**Signal processing for swept frequencies.** Waveforms with the desired $\omega(t)$ characteristics were defined mathematically and loaded to the function generator, with an initial guess for the amplitude envelope versus time based on the known impedance of the RMF coils. The amplitude envelope versus time was then rescaled iteratively based on empirically observed $V_{H_x}$ and $V_{H_y}$ signals to produce a more constant magnitude. The same samples that were used for fixed frequency measurements were also used for swept frequency measurements, with 5 repeated trials interspersed with blanks. Internal oscilloscope averaging was set to 8 for the slow sweeps. Analysis, in brief, consisted of fitting the $V_{H_x}$ curve with the known defined function to verify time delay and correct for minute time scaling artifacts introduced by the oscilloscope (approximately ±2%). With appropriate time coordinate transformation, and with the signal normalized by the known instantaneous frequency, $V_{m_x}$ was used to determine $V_\parallel(t)$ and $V_\perp(t)$ through numerical integration as described in Eq. 5 and 6. $V_\parallel(f)$ and $V_\perp(f)$ were found through the known dependence of $f(t)$. Additional validation of this fitting technique can be found in the *SI Appendix* text and Fig. S3.

**Construction of the magic sphere** The structure designed to hold the magnet array depicted in Fig 4a was 3D printed in 4 parts, with two of these quarters epoxied together to form two halves. Individual elements of permanent ferrite magnets consisted of stacked 20 mm × 20 mm× 3 mm blocks (grade Y35, Supermagnete.ch, FE-Q-20-20-03), forming approximate cubes of 20 mm side length, which were wrapped in electrical tape. The inner equatorial ring of magnets alternated between ferrite and neodymium iron boron with identical geometry (N45, Supermagnete.ch, Q-20-20-03-N). Magnets were loaded into their designated spaces, held by hot glue and epoxy. The two halves were brought together around the inductive detection coil using zip ties, with the zero-point coinciding with the zone for inductive detection. See *SI Appendix*, Fig. S4.

**Selection field experiment.** A three-axis linear micropositioner (SmarAct, controller: HCU-3CX-USB-TAB, piezo positioner: SLC2445 series) was mounted atop the array of permanent magnets and a 3D printed arm was used to hold either the three-axis Hall probe (Metrolab THM1176) or the spherical sample chamber (Fig. 4D, *SI Appendix* Fig. S4). Characterization of the GF and the superposition of the GF and RMF was performed with the Hall probe. Subsequently, the sample chamber was attached to the arm, and the zero point was located where the inductive signal from the micromagnet was maximized. Magnitude was varied at each frequency to empirically maximize out-of-phase signal. Inductive measurements were conducted along the z and y axes at known displacements relative to the zero point, as set by the micro positioner, with blank measurements taken before and after each sweep with the arm removed. In-phase and out-of-phase components were found through direct numerical integration of $V_{m_x}$ with normalized versions of $V_{H_x}$ and $V_{H_y}$ (with any constant offset subtracted and max amplitude set to 1). Blank values subtracted. Resulting out-of-phase signal versus position curves were normalized to their maximum value or to the maximum value projected by a second order spline interpolation in the case of rapidly varying curves.

**Acknowledgments**

M.G.C. was supported by an Eidgenössische Technische Hochschule Postdoctoral Fellowship.

**References**

1. B. J. Nelson, I. K. Kaliakatsos, J. J. Abbott, Microrobots for Minimally Invasive Medicine. *Annual Review of Biomedical Engineering* **12**, 55-85 (2010).
2. J. J. Abbott, E. Diller, A. J. Petruska, Magnetic Methods in Robotics. *Annual Review of Control, Robotics, and Autonomous Systems* **3**, 57-90 (2020).
3. L. Yang, L. Zhang, Motion Control in Magnetic Microrobotics: From Individual and Multiple Robots to Swarms. *Annual Review of Control, Robotics, and Autonomous Systems* **4**, 509-534 (2021).
4. Z. Yang, L. Zhang, Magnetic Actuation Systems for Miniature Robots: A Review. *Advanced Intelligent Systems* **2**, 2000082 (2020).
5. J. J. Abbott *et al.*, How Should Microrobots Swim? *The International Journal of Robotics Research* **28**, 1434-1447 (2009).
6. K. E. Peyer, L. Zhang, B. J. Nelson, Bio-inspired magnetic swimming microrobots for biomedical applications. *Nanoscale* **5**, 1259-1272 (2013).
7. K. E. Peyer, L. Zhang, B. J. Nelson, Localized non-contact manipulation using artificial bacterial flagella. *Applied Physics Letters* **99** (2011).

8. S. Schuerle *et al.*, Synthetic and living micropropellers for convection-enhanced nanoparticle transport. *Science Advances* **5**, eaav4803 (2019).
9. C. Bi, M. Guix, B. V. Johnson, W. Jing, D. J. Cappelleri, Design of Microscale Magnetic Tumbling Robots for Locomotion in Multiple Environments and Complex Terrains. *Micromachines* **9**, 68 (2018).
10. E. E. Niedert *et al.*, A Tumbling Magnetic Microrobot System for Biomedical Applications. *Micromachines* **11**, 861 (2020).
11. Z. Wu *et al.*, Magnetic Mobile Microrobots for Upstream and Downstream Navigation in Biofluids with Variable Flow Rate. *Advanced Intelligent Systems* **4**, 2100266 (2022).
12. Y. Alapan, U. Bozuyuk, P. Erkoc, A. C. Karacakol, M. Sitti, Multifunctional surface microrollers for targeted cargo delivery in physiological blood flow. *Science Robotics* **5**, eaba5726 (2020).
13. U. Bozuyuk *et al.*, Reduced rotational flows enable the translation of surface-rolling microrobots in confined spaces. *Nature Communications* **13**, 6289 (2022).
14. U. Bozuyuk, Y. Alapan, A. Aghakhani, M. Yunusa, M. Sitti, Shape anisotropy-governed locomotion of surface microrollers on vessel-like microtopographies against physiological flows. *Proceedings of the National Academy of Sciences* **118**, e2022090118 (2021).
15. T. O. Tasci, P. S. Herson, K. B. Neeves, D. W. M. Marr, Surface-enabled propulsion and control of colloidal microwheels. *Nature Communications* **7**, 10225 (2016).
16. H. Xie *et al.*, Reconfigurable magnetic microrobot swarm: Multimode transformation, locomotion, and manipulation. *Science Robotics* **4**, eaav8006 (2019).
17. G.-Z. Yang *et al.*, The grand challenges of *Science Robotics*. *Science Robotics* **3**, eaar7650 (2018).
18. C. K. Schmidt, M. Medina-Sánchez, R. J. Edmondson, O. G. Schmidt, Engineering microrobots for targeted cancer therapies from a medical perspective. *Nature Communications* **11**, 5618 (2020).
19. T. Gwisai *et al.*, Magnetic torque-driven living microrobots for increased tumor infiltration. *Science Robotics* **7**, eabo0665 (2022).
20. B. Gleich, J. Weizenecker, Tomographic imaging using the nonlinear response of magnetic particles. *Nature* **435**, 1214-1217 (2005).
21. T. Knopp, T. M. Buzug, *Magnetic particle imaging: an introduction to imaging principles and scanner instrumentation* (Springer Science & Business Media, 2012).
22. A. C. Bakenecker *et al.*, Actuation and visualization of a magnetically coated swimmer with magnetic particle imaging. *Journal of Magnetism and Magnetic Materials* **473**, 495-500 (2019).
23. W. Case, R. Harrington, Calibration of vibrating-sample magnetometers. *J. Res. NBS C* **70**, 225-262 (1966).
24. J. F. Berret, Local viscoelasticity of living cells measured by rotational magnetic spectroscopy. *Nature Communications* **7**, 10134 (2016).
25. L. Chevry, N. K. Sampathkumar, A. Cebers, J. F. Berret, Magnetic wire-based sensors for the microrheology of complex fluids. *Physical Review E* **88**, 062306 (2013).
26. M. A. Rückert *et al.*, Rotational Drift Spectroscopy for Magnetic Particle Ensembles. *IEEE Transactions on Magnetics* **51**, 1-4 (2015).
27. M. Rückert, P. Vogel, V. Behr, Pulsed rotational drift spectroscopy sequences for magnetorelaxometry. *International Journal on Magnetic Particle Imaging IJMPI* **9** (2023).
28. A. Hughes, B. Drury, "Chapter 7 - Variable frequency operation of induction motors" in Electric Motors and Drives (Fifth Edition)*,* A. Hughes, B. Drury, Eds. (Newnes, 2019), https://doi.org/10.1016/B978-0-08-102615-1.00007-6, pp. 229-259.
29. N. Mirkhani, M. G. Christiansen, T. Gwisai, S. Menghini, S. Schuerle, Spatially selective open loop control of magnetic microrobots for drug delivery. *bioRxiv* 10.1101/2023.03.31.535118, 2023.2003.2031.535118 (2023).
30. Y. I. Golovin, A. O. Zhigachev, N. L. Klyachko, D. Y. Golovin, Controlled localization of magnetic nanoparticle mechanical activation in suspension exposed to alternating magnetic field using gradient magnetic field. *Journal of Nanoparticle Research* **24**, 167 (2022).
31. J. Rahmer, C. Stehning, B. Gleich, Spatially selective remote magnetic actuation of identical helical micromachines. *Science Robotics* **2**, eaal2845 (2017).
32. A. Ramos-Sebastian, S.-J. Gwak, S. H. Kim, Multimodal Locomotion and Active Targeted Thermal Control of Magnetic Agents for Biomedical Applications. *Advanced Science* **9**, 2103863 (2022).
33. K. Bente *et al.*, Selective Actuation and Tomographic Imaging of Swarming Magnetite Nanoparticles. *ACS Applied Nano Materials* **4**, 6752-6759 (2021).
34. H. A. Leupold, A. Tilak, E. Potenziani, II, Permanent magnet spheres: Design, construction, and application (invited). *Journal of Applied Physics* **87**, 4730-4734 (2000).
35. P. Blümler, Magnetic Guiding with Permanent Magnets: Concept, Realization and Applications to Nanoparticles and Cells. *Cells* **10**, 2708 (2021).
36. T. F. Sattel *et al.*, Single-sided device for magnetic particle imaging. *Journal of Physics D: Applied Physics* **42**, 022001 (2009).
37. K. Gräfe, A. v. Gladiss, G. Bringout, M. Ahlborg, T. M. Buzug, 2D Images Recorded With a Single-Sided Magnetic Particle Imaging Scanner. *IEEE Transactions on Medical Imaging* **35**, 1056-1065 (2016).

**Main Text Figures**

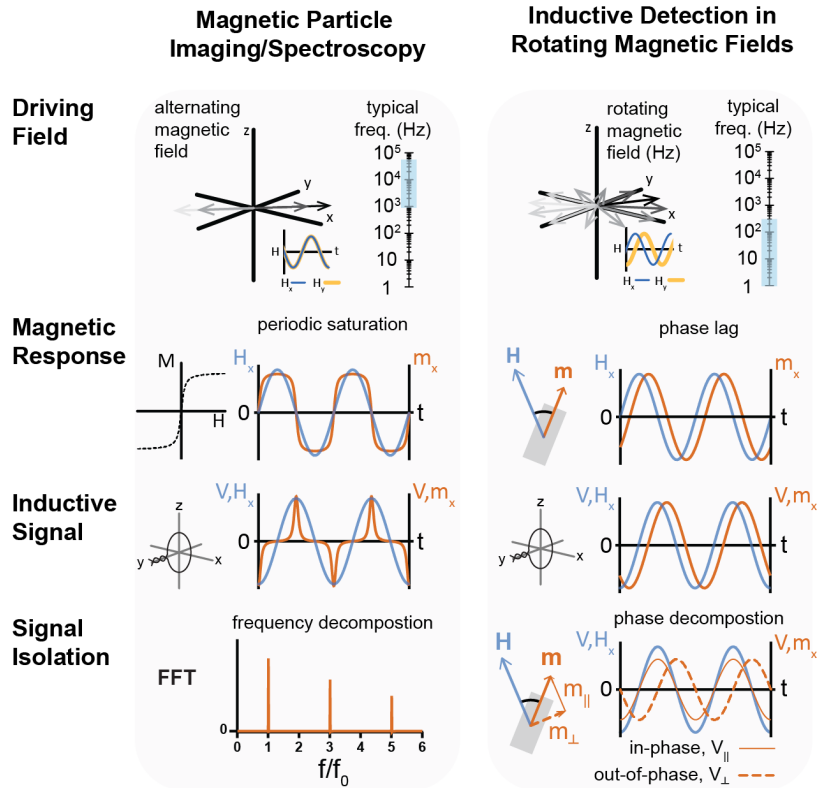

**Figure 1.** Graphical chart comparing salient conceptual aspects of signal acquisition and processing in the familiar case of magnetic particle imaging/spectroscopy with inductive detection of microrobots in rotating magnetic fields (RMFs). The driving field in the case of RMFs has a lower frequency and constant magnitude, leading to magnetic responses without periodic saturation. Consequently, the inductive signals expected from microrobots in RMFs generally do not contain higher order harmonics. Rather than decomposing their signals into different frequency contributions, it is more informative to decompose them in terms of different phase contributions at the same frequency. $H_x$ is the x component of the magnetic field $H$, $m_x$ is the x projection of the magnetic moment $m$, $f$ is frequency, $f_0$ is the principal harmonic in the fast Fourier transform (FFT), $V_{H_x}$ is the voltage induced by $H_x$, and $V_{m_x}$ is the voltage induced by $m_x$. Where applied as subscripts, ∥ and ⊥ denote in-phase or out-of-phase components, respectively.

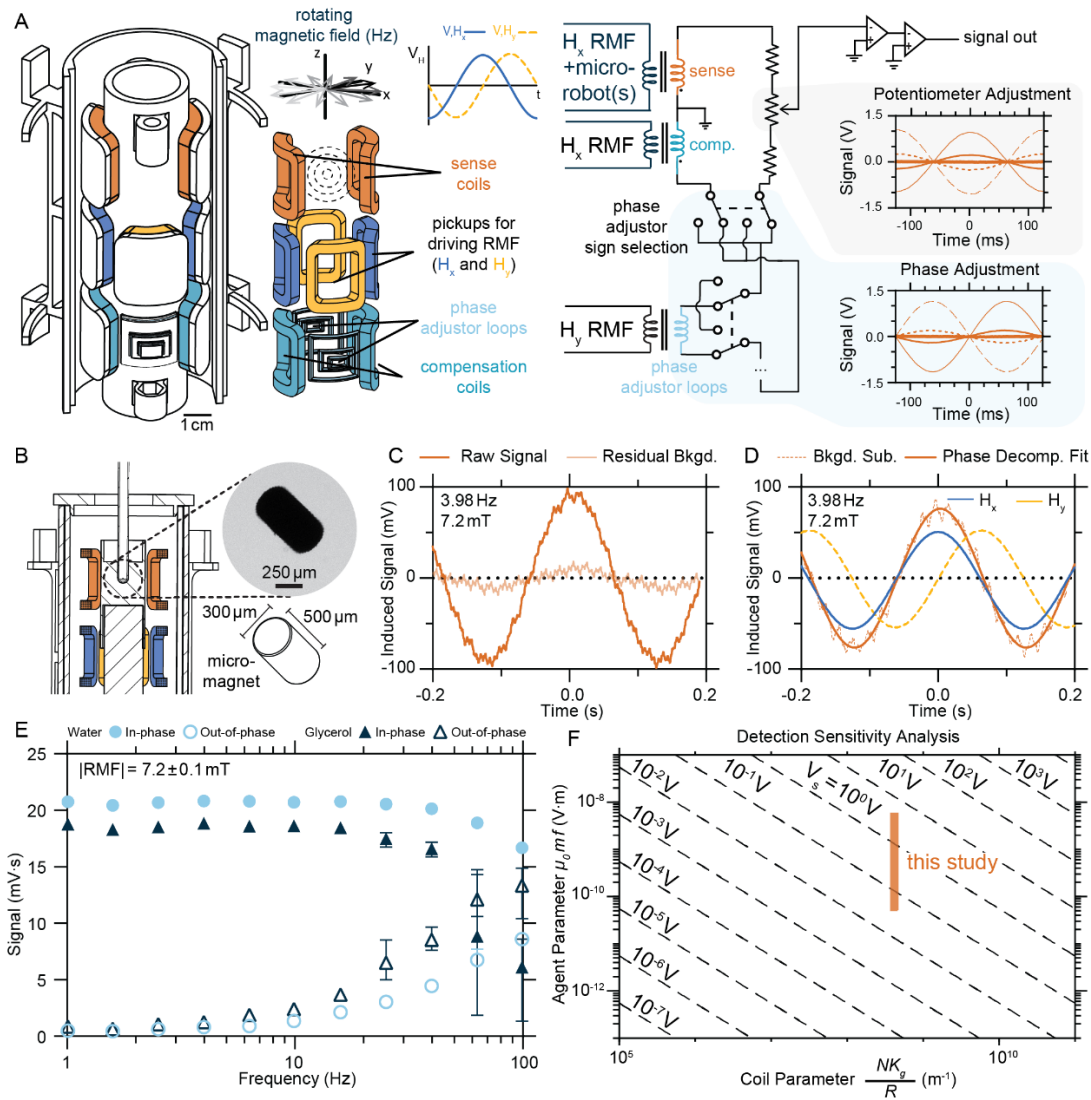

**Figure 2.** Inductive detection of a model microrobot actuated by a rotating magnetic field. (*A*) Sketch of prototype instrument for inductive detection, emphasizing the geometry of the pickup coils. (The coils supplying the RMF are omitted for clarity, but half of the support structure for the 2-phase armature windings is depicted on the left.) Pickup coil pairs for sensing $H_x$ (blue) and $H_y$ (yellow) reside at the center with 1000 turns each, with the sense (orange) and compensation (teal) coil pairs with 5000 turns each appearing above and below. Variably sized phase adjustment loops (light blue) are included near the compensation coils. A simplified electrical schematic of the detection instrument is shown in the center right. For in-phase cancellation adjustment, the signal is taken from the wiper of a potentiometer. A plot of example signals generated by adjusting the potentiometer is provided at right. For out-of-phase cancellation adjustment, phase adjustment loops are selectively incorporated into the circuit with adjustable polarity using the partially represented array of double-pole-double-throw switches, with examples of their contributed out-of-phase signal shown on the bottom right. (*B*) A cross sectional detail of the sample holder situated within the detection apparatus. The model microrobot is a cylindrical micromagnet with a length of 500 µm and a diameter of 300 µm, suspended in water or pure glycerol. Brightfield micrograph of the micromagnet inset. (*C*) Representative example of inductive signal collected for the microrobot in glycerol actuated by a magnetic field of 7.2 mT rotating at 3.98 Hz, with residual signal observed for an identical measurement on a vial containing only glycerol. (*D*) The same signal is shown with the background subtracted, alongside the inductive signals measured by the field pickups. The fit is based on phase decomposition, as described in the text. (*E*) Similar signal collection and phase decomposition was performed over a logarithmic sweep of fixed frequencies for a micromagnet in water and a micromagnet in glycerol. (*F*) A generalized parameter space predicting the scale of the measured induced voltage assuming a geometrically similar detection apparatus, with the range of predicted and observed values for this work highlighted by the orange box. The abscissa is a parameter determined by the radius of the coil $R$, linear gain of the amplifier $K_g$, and number of turns $N$. The ordinate is a parameter set by characteristics of the microrobot, specifically its moment $m$ and frequency of actuation $f$, as well as the permeability of free space $\mu_0$.

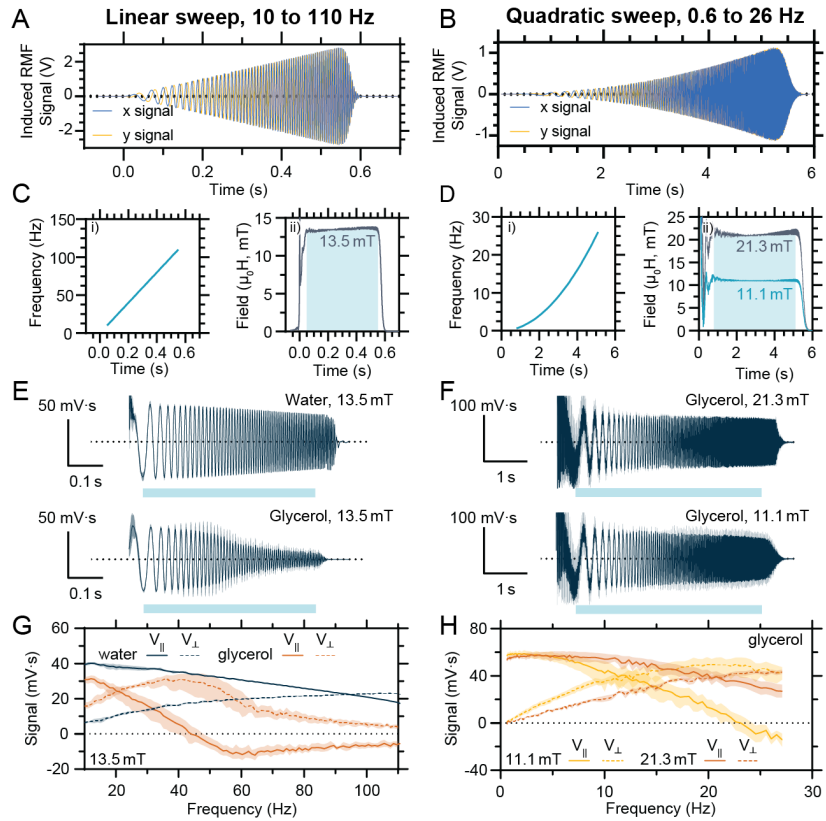

**Figure 3.** Phase decomposition for signals with time-varying frequency. The raw inductive signals generated by the RMF, as detected by the $H_x$ and $H_y$ pickup coils for a rapid linear frequency sweep with constant field magnitude from approximately 10 Hz to 110 Hz (*A*) and a slow quadratic frequency sweep with constant field magnitude from approximately 0.6 Hz to 26 Hz (*B*). Frequency versus time as defined for the input waveforms (i) and field magnitude as inferred from the measured inductive signals as a function of time (ii) are shown for the linear sweep (*C*) and the quadratic sweep (*D*). The timeframe selected for phase decomposition analysis is indicated by the shaded light blue region. Frequency normalized, background subtracted signals are plotted for the same model microrobot as the previous figure under the annotated conditions for the linear sweep (*E*) and quadratic sweep (*F*). The dark blue curve is the mean of 5 replicates, with the shaded bounds representing the 95% confidence interval. The light blue box represents the timeframe selected for decomposition analysis. The signal is decomposed into its instantaneous in-phase component $V_\parallel$ and out-of-phase component $V_\perp$ and plotted as a function of frequency for the linear sweep in (*G*) and quadratic sweep in (*H*). The linear sweep is used to probe the response of nominally identical micromagnets in the distinct viscous environments of water and glycerol. The quadratic sweep is used to probe the variation in the response of a single micromagnet to different driving field amplitudes. Curves represent the mean of 5 replicates, with shaded regions representing the 95% confidence interval.

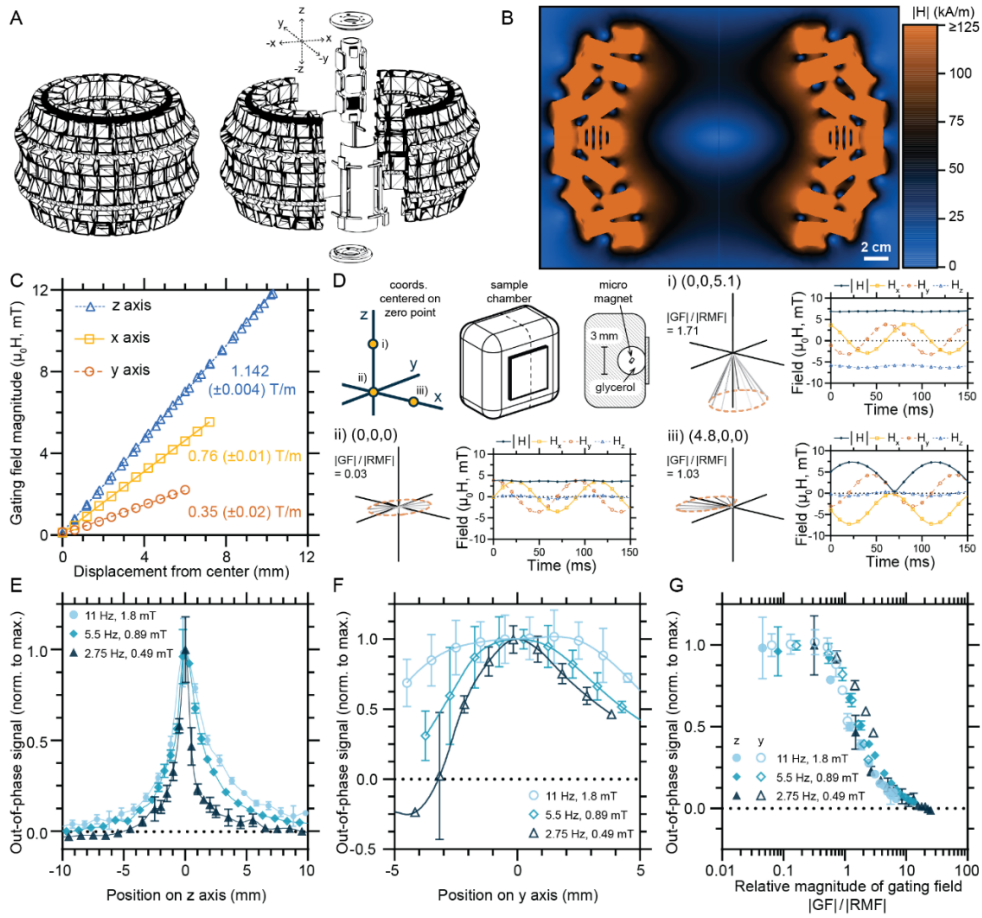

**Figure 4.** Magnetostatic gating fields can spatially isolate inductive signal generation by micromagnets to field free regions when combined with rotating magnetic fields. (*A*) Schematic illustration of the array of permanent magnets used to surround the inductive detection apparatus, a modified "magic sphere." (*B*) Cross sectional representation of a finite element simulation of the GF produced by an azimuthally symmetric approximation of the permanent magnet array. Made with FEMM. (*C*) Measured magnetostatic field magnitude as a function of displacement from the zero point along each of the indicated axes. Slopes extracted from separate linear regressions are shown with 95% confidence interval bounds. (*D*) Sketch of the sample chamber used to test inductive signal generation within a magnetostatic GF as a function of position, consisting of a 3 mm spherical cavity filled with glycerol and containing a model cylindrical micromagnet (500 μm length, 300 μm diameter). Results of characterization of the net field produced by the combined GF and RMF are shown at several representative points: i) displaced along the positive z direction, ii) at the zero point, and iii) displaced along the positive x direction. Curves are sinusoidal fits of data taken with a Hall probe. (*E*) and (*F*) show the variation of the out-of-phase component of the inductive signal produced by the model micromagnet with position along the z axis and y axis, respectively. Each curve represents measurements under distinct RMF conditions (1.84 mT at 11 Hz, 0.89 mT at 5.5 Hz, and 0.49 mT at 2.75 Hz) and is normalized to the maximum mean out-of-phase signal observed or interpolated by a second order spline curve. Error bars represent 95% confidence interval, with N = 3 replicates. (*G*) Relative magnitude of the GF, defined as the quotient of the GF magnitude experienced at a point to the magnitude of the superimposed RMF, is used to replot the normalized out-of-phase signal data from positive displacements in (*E*) and (*F*).

**Supporting Information for**

Inductive sensing of magnetic microrobots under actuation by rotating magnetic fields

Michael G. Christiansen*[1], Lucien Stöcklin[1,2], Cameron Forbrigger[1], Shashaank Abhinav Venkatesh [1,3], Simone Schuerle*[1]

1 Department of Health Sciences and Technology, ETH Zurich
2 Department of Biosystems Science and Engineering, ETH Zurich
3 Department of Biomedical Engineering, National University of Singapore

*Michael G. Christiansen and *Simone Schuerle

**Email:** michael.christiansen@hest.ethz.ch, simone.schuerle@hest.ethz.ch

**Supporting Information Text**

**Theoretical sensitivity analysis.** Here, details are provided to substantiate the sensitivity analysis conducted for Fig. 2F of the main text. For simplicity, the geometry of the sense coils, which are partial cylindrical surfaces (Fig. 2A), are treated here approximately as partial spherical surfaces. Specifically, they are treated as latitude-longitude squares on the surface of a sphere of radius $R$ with polar angle limits $\theta_1 \leq \theta \leq \theta_2$ and azimuth angle limits $\varphi_1 \leq \varphi \leq \varphi_2$ (Fig. S2).

Given a single loop of conducting wire that encloses a surface $S$, the voltage $V$ induced in the loop by a magnetic field $\boldsymbol{B}$ is given by Faraday's law of electromagnetic induction:

$$V(t) = \frac{d}{dt}\left(\int_S \boldsymbol{B} \cdot \hat{\boldsymbol{n}}\, dA\right) \qquad [1]$$

where $t$ is time, $dA$ is a differential area element of the surface $S$, and $\hat{\boldsymbol{n}}$ is the normal vector to that differential area element. A coil composed of $N$ perfectly overlapping loops results in $N$ times higher induced voltage.

Given a magnetic point dipole at a position $\boldsymbol{P}_m$, the magnetic field from the magnetic dipole can be determined analytically for any position of interest $\boldsymbol{P}$. The equation that describes the magnetic field from a magnetic point dipole in 3D space as a function of the relative position $\boldsymbol{r} = \boldsymbol{P} - \boldsymbol{P}_m$ is:

$$\boldsymbol{B}_m = \frac{\mu_0 m}{4\pi \|\boldsymbol{r}\|^3}(3\hat{\boldsymbol{r}}\hat{\boldsymbol{r}}^{\mathrm{T}} - I_3)\widehat{\boldsymbol{m}} \qquad [2]$$

where $\mu_0$ is the permeability of free space, $\|\boldsymbol{r}\|$ is the 2-norm of $\boldsymbol{r}$, $\hat{\boldsymbol{r}} = \boldsymbol{r}/\|\boldsymbol{r}\|$ is a unit vector in the direction of $\boldsymbol{r}$, and $I_3$ is a 3 × 3 identity matrix (1).

For a dipole rotating within the xy plane at a constant angular velocity $\omega$, the resulting magnetic field expressed in spherical coordinates becomes

$$\boldsymbol{B}_m(t) = \frac{\mu_0 m}{4\pi r^3}\begin{bmatrix} 2\sin\theta \cos(\omega t - \varphi) \\ -\cos\theta \cos(\omega t - \varphi) \\ -\sin(\omega t - \varphi) \end{bmatrix} \qquad [3]$$

Substituting Eq. 3 into Eq. 1 in order find an estimate for the induced voltage contribution of the dipole $V_m$,

$$V_m(t) = \frac{\mu_0 m}{2\pi R}\frac{d}{dt}\left(\int_{\varphi_1}^{\varphi_2}\int_{\theta_1}^{\theta_2} \cos(\omega t - \varphi)\sin^2\theta\, d\theta d\varphi\right) \qquad [4]$$

For this simplified case, the solution is analytically tractable, and can be expressed in terms of the following parameters:

$$\theta_\Delta = \frac{\theta_2 - \theta_1}{2} \qquad [5]$$

$$\bar{\theta} = \frac{\theta_1 + \theta_2}{2} \qquad [6]$$

$$A_\theta = 2\theta_\Delta - \sin(2\theta_\Delta)\cos(2\bar{\theta}) \qquad [7]$$

$$\varphi_\Delta = \frac{\varphi_2 - \varphi_1}{2} \tag{8}$$

$$\bar{\varphi} = \frac{\varphi_1 + \varphi_2}{2} \tag{9}$$

$$A_\varphi = \sin(\varphi_\Delta) \tag{10}$$

The full expression for $V_m(t)$ can then be given as

$$V_m(t) = -\frac{\mu_0 m \omega}{2\pi R} A_\theta A_\varphi \sin(\omega t - \bar{\varphi}) \tag{11}$$

The parameters $A_\theta$ and $A_\varphi$ are unitless geometric factors; therefore we can see that the voltage induced by a rotating dipole scales with $1/L$, where $L$ is the characteristic length of the coils. As expected, the induced voltage scales linearly with respect to the dipole magnitude $m$ and the rotation frequency ω. The average azimuth angle $\bar{\varphi}$ of the coil only results in a phase shift of the induced voltage signal and has no effect on the magnitude of the signal. The effect of $\varphi_\Delta$ is also relatively simple: $A_\varphi$ is maximized at $\varphi_\Delta = \pi/2$, $A_\varphi = 1$ (when the width of the coil spans a full 180°) and decreases symmetrically about that maximum.

The method of physical background subtraction shown in Fig. 2A, in which the voltage induced by the rotating magnetic moment is ultimately taken from the wiper of the potentiometer at the center of a voltage divider, results in the magnitude of the sensed voltage dropping by a factor of two. The voltage that is sensed by the oscilloscope is also amplified, increasing the scale of the measured voltage by a factor of $K_g$, where $K_g$ is the linear gain of the amplifier. By grouping the relevant parameters determined by the coil setup (coil parameter) and constrained by the microrobot (agent parameter) the expected scale of measured voltages can be mapped over many orders of magnitude (main text, Fig. 2F).

It is enlightening to further consider the scale of the voltages induced in the sense coils by the RMF itself. This enables both a comparison between the magnitude of the voltage induced by the RMF and the microrobot ("signal to background ratio" $SBR$) as well as the extent of background suppression that was ultimately achieved with our setup ("background rejection ratio" $BRR$). In spherical coordinates, a field with the same plane of rotation as the one assumed for m and also rotating at constant angular velocity $\omega$ can be expressed as follows:

$$\boldsymbol{B}_{RMF}(t) = B_{RMF} \begin{bmatrix} \sin\theta \cos(\omega t - \varphi) \\ \cos\theta \cos(\omega t - \varphi) \\ \sin(\omega t - \varphi) \end{bmatrix} \tag{12}$$

Substituting this expression into Eq. 1 yields

$$V_{RMF}(t) = -B_{RMF} R^2 \omega A_\theta A_\varphi \sin(\omega t - \bar{\varphi}) \tag{13}$$

Comparing Eq. 11 and 13, the time dependent part of this function and the geometric factor are identical and will drop out of a ratio, although it should be noted that a real microrobot would experience a phase lag between field and the moment for reasons discussed in the main text. Dividing Eq. 11 by Eq. 13 gives the $SBR$,

$$SBR = \frac{\mu_0 m}{2\pi B R^3} \qquad [14]$$

For the microrobots used in this study, $m$ is approximately 4.12×10$^{-5}$ Am$^{-2}$, $R$ was approximately 19.5 mm, and the $SBR$ evaluates to values on the order of 10$^{-4}$ to 10$^{-3}$ for field magnitudes in the range of 1 to 10 mT.

The *BRR* for the representative traces in Fig. 2C can also be estimated. The residual background at 3.98 Hz and 7.2 mT observed for the blank sample can be estimated through fitting to have an amplitude of 9.725 mV. At this frequency, the characterized gain of the amplifier $K_g$ was about 3255, implying the actual residual voltage was about 2.988 µV. For the geometry and turn number of turns present in the sense coil, $\theta_\Delta \approx \varphi_\Delta \approx \pi/4$, $N = 5000$, and $R \approx 19.5$ mm, giving an estimated voltage within the sense coil from the stated RMF of 138.2 mV. This suggests a $BRR$ value of $4.62 \times 10^4$, or 93.3 dB. Since the residual signal and the uncompensated voltages occurring in the sense coils should scale similarly with $B_{RMF}$ and $\omega$, the $BRR$ should remain about the same for all conditions investigated, while the apparent magnitude of the residual voltage increases. This analysis suggests that, after background had been properly adjusted, a background suppression of about 90 dB could be achieved with this setup.

**Validation of frequency sweep phase decomposition analysis.** Some additional details can be provided regarding phase decomposition with the frequency sweeps in the main text. For the case of a linear sweep with a constant rate $\alpha$,

$$\omega(t) = \alpha t \qquad [15]$$

Here, the weighting function $W(t)$ is as follows

$$W(t) = \frac{d}{dt}[\omega(t)t] = \frac{d}{dt}(\alpha t^2) = 2\alpha t \qquad [16]$$

Advancing a single period after an initial time $t_i$ to a final time $t_f$, it is possible to find an expression for $t_f$ in terms of $t_i$ and $\alpha$

$$\alpha t_f^2 = \alpha t_i^2 + 2\pi \qquad [17]$$

$$t_f = \sqrt{t_i^2 + \frac{2\pi}{\alpha}} \qquad [18]$$

The angular frequency corresponding to these limits can be taken as the time average $\langle\omega(t)\rangle$ over a single period. Here,

$$\langle\omega(t)\rangle = \frac{1}{t_f - t_i}\int_{t_i}^{t_f}\omega(t) = \frac{1}{2}\frac{\alpha(t_f^2 - t_i^2)}{2\pi(t_f - t_i)} = \frac{\alpha(t_f - t_i)(t_f + t_i)}{4\pi(t_f - t_i)} = \frac{\alpha(t_f + t_i)}{4\pi} \qquad [19]$$

In the case of the linear sweep results shown in the main text, $\alpha = 400\pi$. To ensure a smooth rise and fall of the field, sigmoidal step functions are introduced such that the full expression for the desired field as a function of time is as follows:

$$H_x(t) = H_0 \cos(400\pi t^2)\left\{\frac{1}{1 + \exp[-150(t - 0.025)]} - \frac{1}{1 + \exp[-150(t - 0.575)]}\right\} \qquad [20]$$

$$H_y(t) = H_0 \sin(400\pi t^2)\left\{\frac{1}{1 + \exp[-150(t - 0.025)]} - \frac{1}{1 + \exp[-150(t - 0.575)]}\right\} \qquad [21]$$

These functions, as well as their modulus, are plotted in Fig. S3A. The time of constant field magnitude is approximately [0.05,0.55]. A more important constraint for performing phase decomposition with the approximate forms in Eq. 5 and 6 from the main text is to verify orthogonality of the basis functions. In other words, for Eq. 5 and 6 to describe a valid approximation, the following should be true for each set of suitable limits:

$$\frac{2\int_{t_i}^{t_f} W(t)\cos^2[\omega(t)t]\,dt}{\omega(t_f)t_f - \omega(t_i)t_i} \approx \frac{2\int_{t_i}^{t_f} W(t)[H_x(t)/H_0]^2\,dt}{\omega(t_f)t_f - \omega(t_i)t_i} \approx 1 \quad [22]$$

$$\frac{2\int_{t_i}^{t_f} W(t)\sin^2[\omega(t)t]\,dt}{\omega(t_f)t_f - \omega(t_i)t_i} \approx \frac{2\int_{t_i}^{t_f} W(t)[H_y(t)/H_0]^2\,dt}{\omega(t_f)t_f - \omega(t_i)t_i} \approx 1 \quad [23]$$

$$\frac{2\int_{t_i}^{t_f} W(t)\sin[\omega(t)t]\cos[\omega(t)t]\,dt}{\omega(t_f)t_f - \omega(t_i)t_i} \approx \frac{2\int_{t_i}^{t_f} W(t)[H_x(t)H_y(t)/H_0^2]\,dt}{\omega(t_f)t_f - \omega(t_i)t_i} \approx 0 \quad [24]$$

For the case of the linear sweep, good convergence to these expected values is observed (Fig S3B). The absolute value of the error function is also plotted to quantify deviation from orthogonality (Fig. S3C). All points where the error falls below 0.01 were taken to be valid and included later in the analysis of actual signals. At this stage the deviation is purely numerical—this does not include the influence of noise or instrumental error in actual signals. The symmetry of the error function suggests that, in this case, the main source of error may be the residual influence of the sigmoidal step functions.

This process can also be repeated explicitly for a quadratic sweep with prefactor $\beta$:

$$\omega(t) = \beta t^2 \quad [25]$$

In this case,

$$W(t) = \frac{d}{dt}[\omega(t)t] = \frac{d}{dt}(\beta t^3) = 3\beta t^2 \quad [26]$$

The expression relating the limits $t_i$ and $t_f$ reduces to:

$$\beta t_f^3 = \beta t_i^3 + 2\pi \quad [27]$$

$$t_f = \left(t_i^3 + \frac{2\pi}{\beta}\right)^{1/3} \quad [28]$$

Once again, the angular frequency these limits correspond to can be found using the time averaged value of $\omega(t)$:

$$\langle\omega(t)\rangle = \frac{1}{t_f - t_i}\int_{t_i}^{t_f} \frac{\omega(t)}{2\pi} = \frac{1}{6\pi}\frac{\beta(t_f^3 - t_i^3)}{t_f - t_i} = \frac{\beta(t_f - t_i)(t_f^2 + t_f t_i + t_i^2)}{6\pi(t_f - t_i)}$$
$$= \frac{\beta}{6\pi}(t_f^2 + t_f t_i + t_i^2) \qquad [29]$$

In the case of the quadratic sweep in the main text, $\beta = 2\pi$ and the time of constant field is from approximately [0.5,5.5]. The explicit forms of $H_x(t)$ and $H_y(t)$ are as follows:

$$H_x(t) = H_0 \cos(2\pi t^3)\left\{\frac{1}{1 + \exp[-15(t - 0.5)]} - \frac{1}{1 + \exp[-15(t - 5.5)]}\right\} \qquad [30]$$

$$H_y(t) = H_0 \sin(2\pi t^3)\left\{\frac{1}{1 + \exp[-15(t - 0.5)]} - \frac{1}{1 + \exp[-15(t - 5.5)]}\right\} \qquad [31]$$

These are shown in Fig. S3D. The same check on orthogonality described in Eq. 22-24 can be performed (Fig. S3E and Fig S3F). Here, the errors are larger than in the linear case, but convergence falls below an error of 0.01 for points that were considered in the analysis.

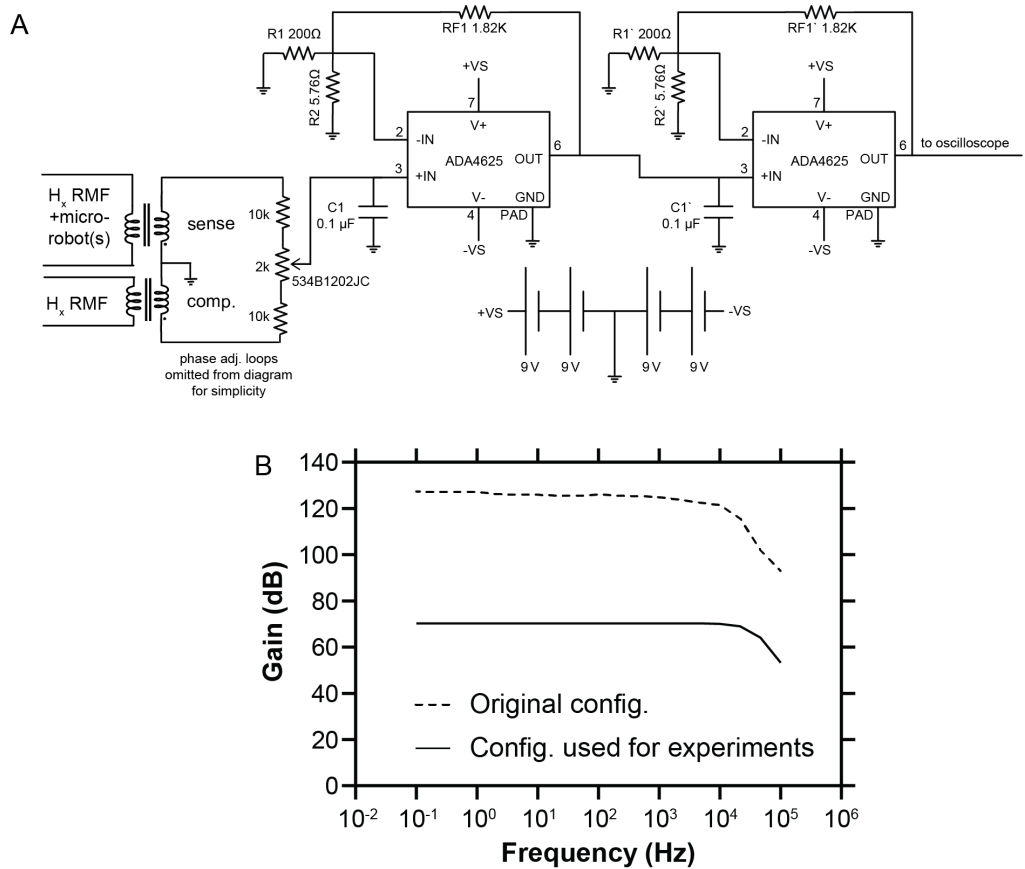

**Fig. S1.** Additional information about signal amplification in the experiments described in the main text. (*A*) A more detailed schematic of the amplifier circuit is shown. For PCB layout, two of the evaluation boards EVAL-ADA4625-1ARDZ were used with the component values indicated. C1 was found to be necessary to suppress self-resonance of the detection coils. Gain was set by changing RF1 and R2, or RF1' and R2'. The signal amplifier was powered by standard 9 V batteries to fully avoid the possibility for noise from a rectified DC supply. (*B*) Measured gain versus frequency is shown for an initial configuration of the amplifier, which achieved a gain of more than 120 dB, as well as the gain setting used for experiments (70 dB). The gain was reduced so that the micromagnet could be observed at the highest frequencies. It was also beneficial to reduce the gain because additional suppression of the ambient 50 Hz noise is needed to avoid clipping output waveforms.

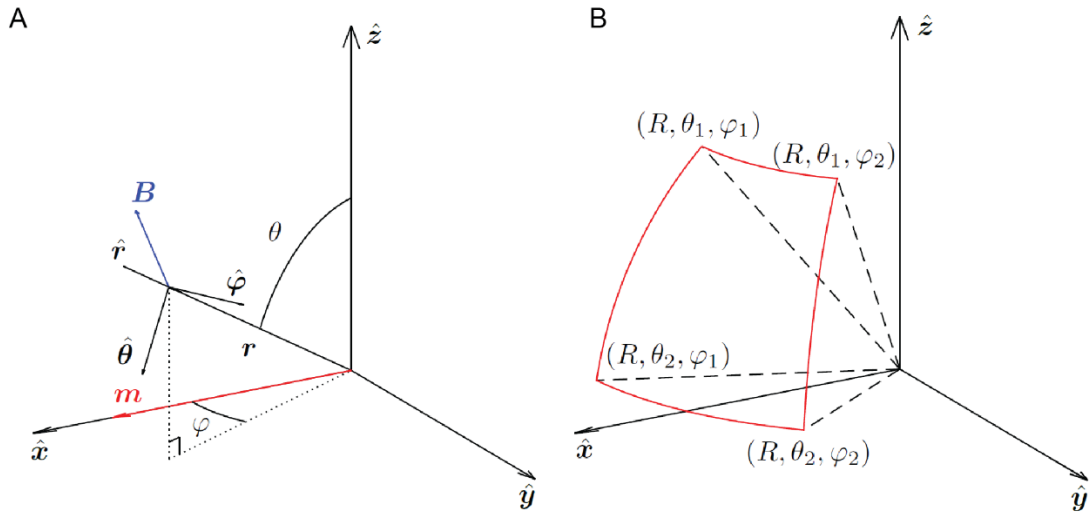

**Fig. S2.** Explanatory sketch of geometry for theoretical sensitivity analysis. (*A*) The magnetic field vector $B$ at a position $r$ relative to a point dipole with dipole moment vector $m$. Both Cartesian $\hat{x}\hat{y}\hat{z}$ and spherical $\hat{r}\hat{\theta}\hat{\varphi}$ coordinate frames are shown. The Cartesian frame and the spherical frame have the same origin, but the spherical frame vectors are drawn at the tip of $r$ to allow for clearer visualization. (*B*) A loop of wire that forms a latitude-longitude rectangle $\theta_1 \leq \theta \leq \theta_2$, $\varphi_1 \leq \varphi \leq \varphi_2$ on the surface of a sphere with radius $R$ centered at the origin. Note that the coils actually used in the apparatus described in the main text were partial cylindrical surfaces, which here have been approximated in terms of latitude longitude squares to simplify the analytical problem.

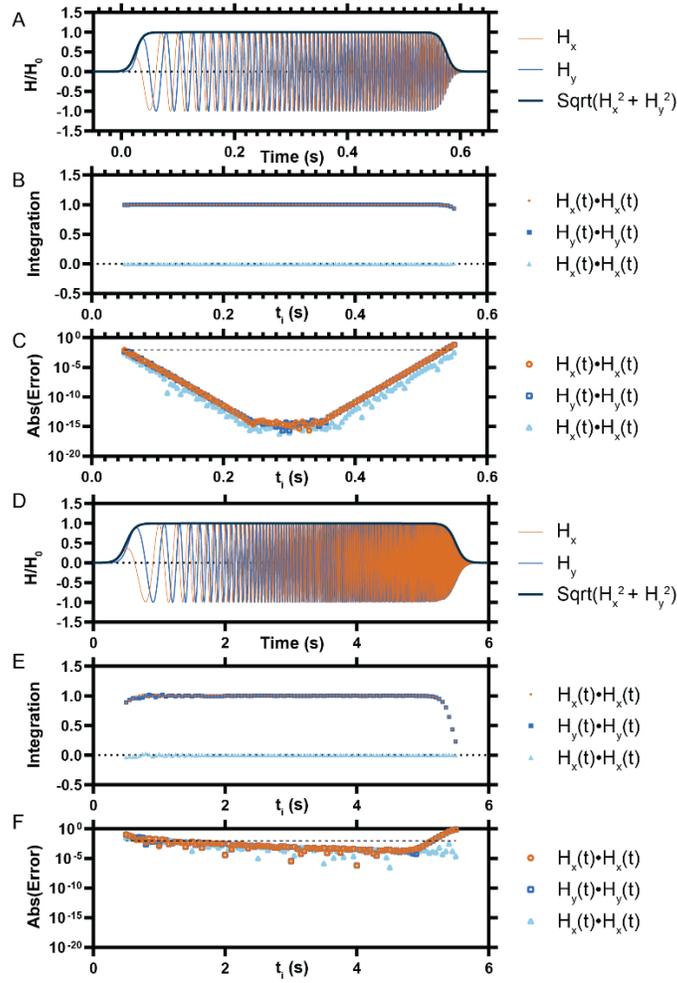

**Fig. S3.** Numerical validation of phase decomposition with swept frequencies for the cases shown in the main text. Panels (*A*)-(*C*) pertain to the linear sweep from approximately 10 to 100 Hz described in Fig. 3A of the main text. (*A*) The desired field as a function of time, normalized to the RMF magnitude $H_0$ is shown in terms of its components $H_x$ and $H_y$. (*B*) The orthogonality of these functions when integrated with a weighting function and appropriately defined limits is tested by assessing convergence to 1 or 0. As simplified notation, $H_x(t) \cdot H_x(t)$ is used to denote results from Eq. 22, $H_y(t) \cdot H_y(t)$ for Eq. 23, and $H_x(t) \cdot H_y(t)$ for Eq. 24. (*C*) To quantify the level of convergence, the absolute value of the difference between calculated and expected values is shown. Panels (*D*)-(*F*) pertain to the quadratic sweep from approximately 0.6 Hz to 26 Hz described in Fig. 3B of the main text and are directly analogous to (*A*)-(*C*). It should be noted in (*F*) that the error is higher in the case of the quadratic sweep, but still acceptably low for the timeframe of interest.

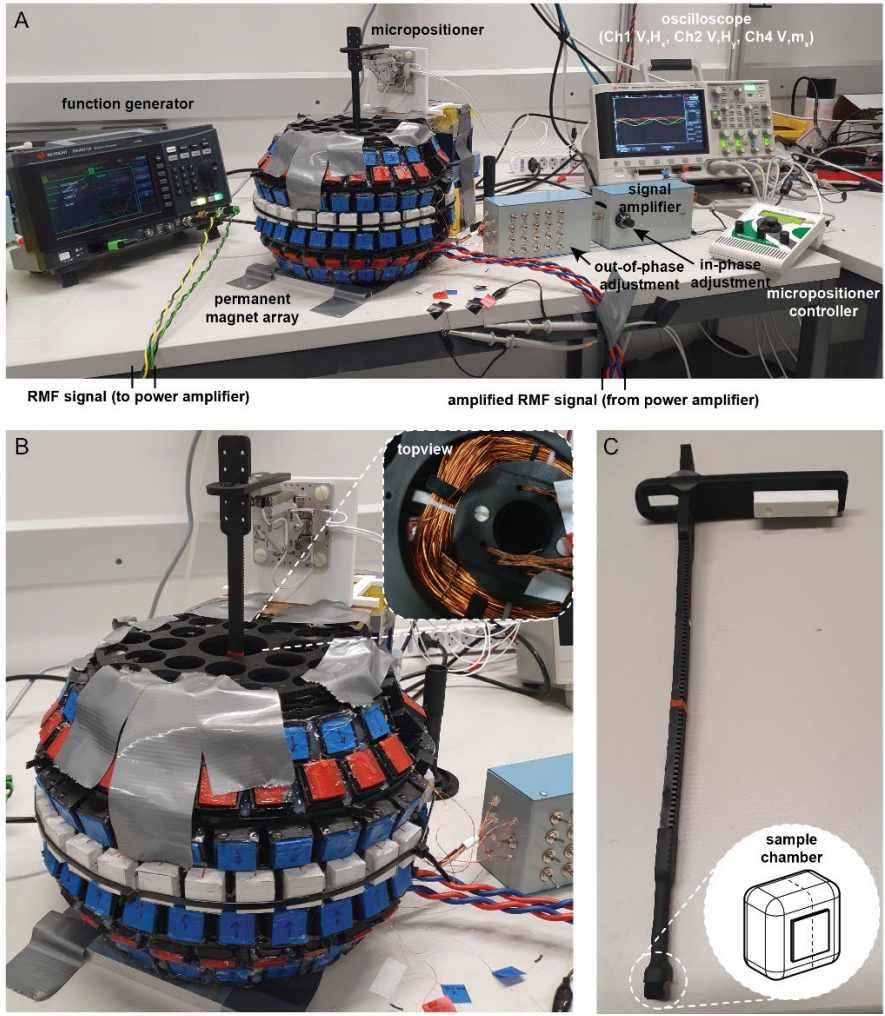

**Fig. S4.** Images of the selection field setup. (*A*) The full setup is depicted as actually used to collect the data appearing in Fig. 4 of the main text, with the components labelled. The power amplifier is not included because it was relocated across the room to reduce 50 Hz noise, but its inputs and outputs are labelled. (*B*) A detailed view of the permanent magnet array used to generate the zero point, along with the micro positioner mounted atop it. A topview is provided with the arm holding the sample removed (inset). (*C*) The arm that holds the sample chamber and attaches to the micropositioner stage is shown. The sample chamber is affixed to the arm using a piece of heat shrink.

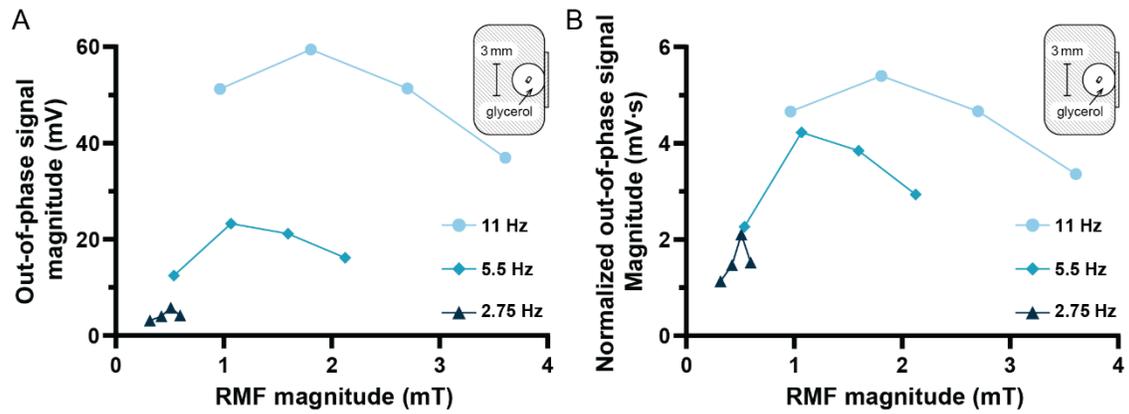

**Fig. S5.** Empirical determination of RMF conditions suitable for the selection field experiments. (*A*) Out-of-phase signal amplitudes are shown as a function of RMF magnitude for the sample chamber placed as close to the zero point as possible. As expected, the out-of-phase signal increases with RMF magnitude, but then drops again as the signal from the magnet is dominated by the in-phase component. As discussed in the main text, these signals are proportional to the rate of irreversible work done on the micromagnet, so the peak corresponds roughly to the step out frequency. (*B*) By normalizing the same data to frequency, it can be replotted in terms of quantities that reflect the magnitude of the irreversible torque or the irreversible work per cycle of the field.

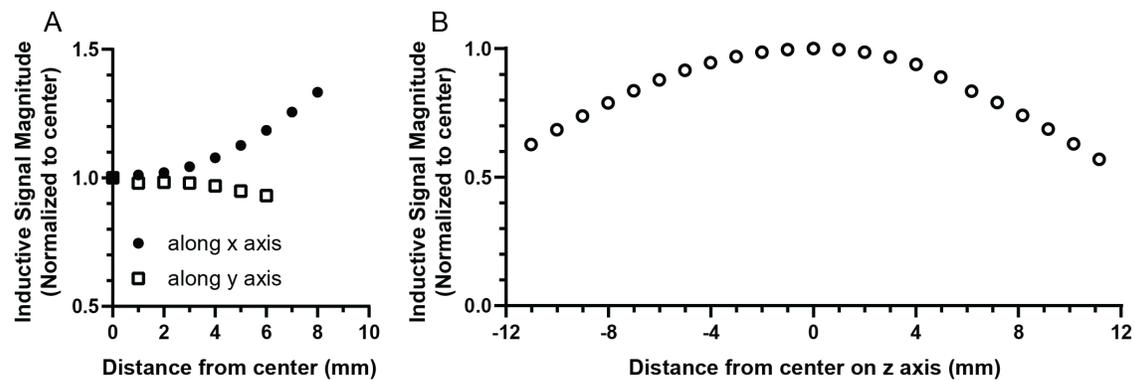

**Fig. S6.** Assessment of spatial variation of inductive coupling in absence of selection field. To ensure a large signal at low frequency, a cylindrical magnet of diameter 750 μm and length of 1000 μm (N50, SM Magnetics Cyl0010-25) was placed in glycerol in a 5 mm NMR tube and its position was varied by the microcontroller under constant RMF. Variation in magnitude of the induced signal under these conditions was assumed to arise from spatial variation of inductive coupling to the sense coils. (*A*) Variation in signal magnitude along the x and y axes is shown, normalized to the center. Notably, the observed inductive coupling is more constant for variation in the y direction than the x direction for this setup. (*B*) Variation of the inductive signal magnitude is shown in the z direction. To facilitate comparison, the ordinate axes of these plots have been kept to scale with each other. Variation of inductive coupling in the z direction is clearly much weaker than the influence of the selection field.

**SI References**

1. J. J. Abbott, E. Diller, A. J. Petruska, Magnetic Methods in Robotics. *Annual Review of Control, Robotics, and Autonomous Systems* **3**, 57-90 (2020).